\documentclass[journal=apchd5,manuscript=article]{achemso}
\usepackage{epstopdf}
\usepackage[]{textcomp}
\usepackage[]{xspace}
\usepackage{amsmath}
\usepackage{color}

\title{Ultrafast Plasmonic Control of Second Harmonic Generation}

\author{Roderick B. Davidson II}
\email{roderick.b.davidson@vanderbilt.edu}
\affiliation{Department of Physics and Astronomy, Vanderbilt University, Nashville, Tennessee 37235-1807, USA}
\alsoaffiliation{Computational Sciences and Engineering Division, Oak Ridge National Laboratory, Oak Ridge, Tennessee 37831, USA}
\author{Anna Yanchenko}
\affiliation{Department of Physics and Astronomy, Vanderbilt University, Nashville, Tennessee 37235-1807, USA}
\alsoaffiliation{Department of Physics, University of Virginia, Charlottesville, Virginia 22904-4714, USA}
\author{Jed I. Ziegler}
\author{Sergey M. Avanesyan}
\affiliation{Department of Physics and Astronomy, Vanderbilt University, Nashville, Tennessee 37235-1807, USA}
\author{Ben J. Lawrie}
\affiliation{Computational Sciences and Engineering Division, Oak Ridge National Laboratory, Oak Ridge, Tennessee 37831, USA}
\author{Richard F. Haglund Jr.}
\affiliation{Department of Physics and Astronomy, Vanderbilt University, Nashville, Tennessee 37235-1807, USA}

\begin{document}

\begin{abstract}  Efficient frequency conversion techniques are crucial to the development of plasmonic metasurfaces for information processing and signal modulation. In principle, nanoscale electric-field confinement in nonlinear materials enables higher harmonic conversion efficiencies per unit volume than those attainable in bulk materials. Here we demonstrate efficient second-harmonic generation (SHG) in a serrated nanogap plasmonic geometry that generates steep electric field gradients on a dielectric metasurface. An ultrafast pump is used to control plasmon-induced electric fields in a thin-film material with inversion symmetry that, without plasmonic enhancement, does not exhibit an an even-order nonlinear optical response. The temporal evolution of the plasmonic near-field is characterized with $\sim100$~as resolution using a novel nonlinear interferometric technique. The ability to manipulate nonlinear signals in a metamaterial geometry as demonstrated here is indispensable both to understanding the ultrafast nonlinear response of nanoscale materials, and to producing active, optically reconfigurable plasmonic devices.\end{abstract}

The manipulation of nonlinear optical phenomena using plasmonic control and near-field enhancement has recently enabled new ways to create active nanoscale photonic elements.\cite{kauranen2012nonlinear,hess2012active,melikyan2014high} Several geometries have been used in investigations of the effects of plasmonic field confinement and asymmetry in second-order nonlinear optical processes.\cite{davidson2015efficient,aouani2012multiresonant,zhang2011three,capretti2012multipolar,capretti2014size} In addition, dielectric materials near the surface of plasmonic nanostructures also exhibit enhanced nonlinear optical responses proportional to the confinement of the electric field.\cite{boyd2003nonlinear,zernike2006applied, lee2014giant, aouani2014third} The short excitation lifetime and nanoscale footprint of metasurface devices hold promise for replacing macroscopic nonlinear crystals in biophotonic applications and tunable light sources. While the efficiency of these coupled metal-dielectric systems continues to improve, controlling these systems at timescales that would utilize the ultrashort plasmonic lifetimes remains an obstacle to fabricating nonlinear plasmonic devices that can be used for active manipulation of optical signals. Current plasmonic technology has yet to deliver a method of characterizing and controlling nonlinear signals on the ultrafast timescale.

Although active control over nonlinear plasmonic interactions has been achieved by applying DC electric fields\cite{cai2011electrically,kang2014electrifying} the size of contacts and the frequency response of electrical excitation defeat the principal advantages afforded by plasmonic systems, most notably nanoscale footprints and femtosecond interaction times. Here we demonstrate plasmonically controlled second-harmonic generation mediated by optically induced plasmon fields in a metal-dielectric nanocomposite, based on a model platform that also enables the direct characterization of nanoscale dielectric materials exposed to electric fields oscillating at optical frequencies. Second-harmonic interferometric spectroscopy performed on metal-dielectric nanocomposite systems allows us to map the dynamics of plasmonic oscillations on a sub-femtosecond timescale and clearly distinguish the SHG contributions from the enhanced dielectric from that of plasmonic scattering.

\begin{figure*}[!t]
\includegraphics[width=\columnwidth]{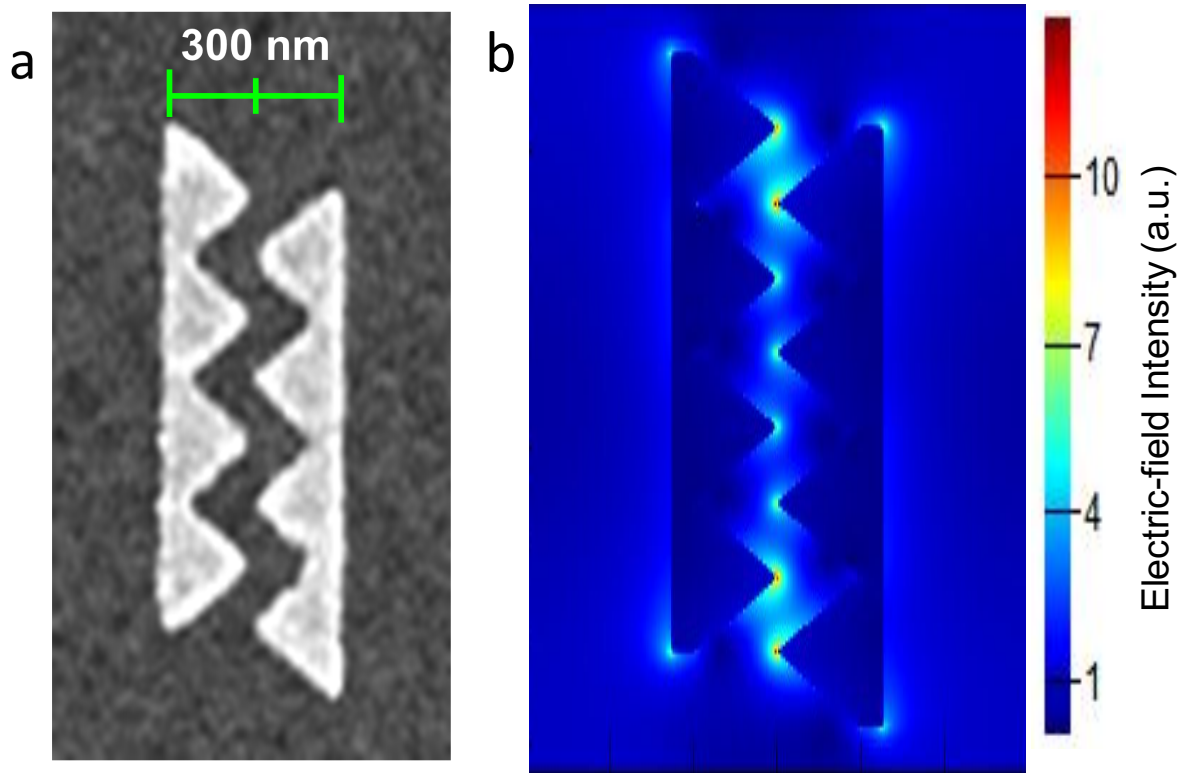}
\caption[fig1]{\label{fig:fig1}\textbf{The serrated nanogap. a,} Scanning-electron microscope image of a single element of the SNG array \textbf{b,} Finite-difference, time-domain simulation of the near-field intensity of the SNG plasmon excited by a pulse polarized perpendicular to the nanogap or horizontal to the image at $\lambda$ = 800nm.}
\end{figure*}

The array of lithographically fabricated gold serrated nanogaps (SNGs), shown in Fig.1, create a field of electric-field hotspots in a planar, capacitor-like geometry. When the nanostructure is excited by a control laser pulse (800~nm wavelength, 30~fs pulse duration) polarized perpendicular to the gap, regions of high electric field form at the points of the teeth, creating steep electric field gradients within the nanogap. Fig. 1b shows finite-difference, time-domain (FDTD) simulations of the plasmonic evanescent electric field intensity generated by the control pulse. When the nanogap is filled by spin-coating poly(methyl methacrylate) (PMMA) onto the sample and the plasmon is excited using an ultrafast laser pulse, intense electric fields within the dielectric oscillate with a period of approximately 2.7~fs. These oscillations rapidly polarize the electronic structure of the PMMA, causing large, time-dependent changes in its effective second-order susceptibilities.

\begin{figure*}[!t]
\includegraphics[width=\textwidth]{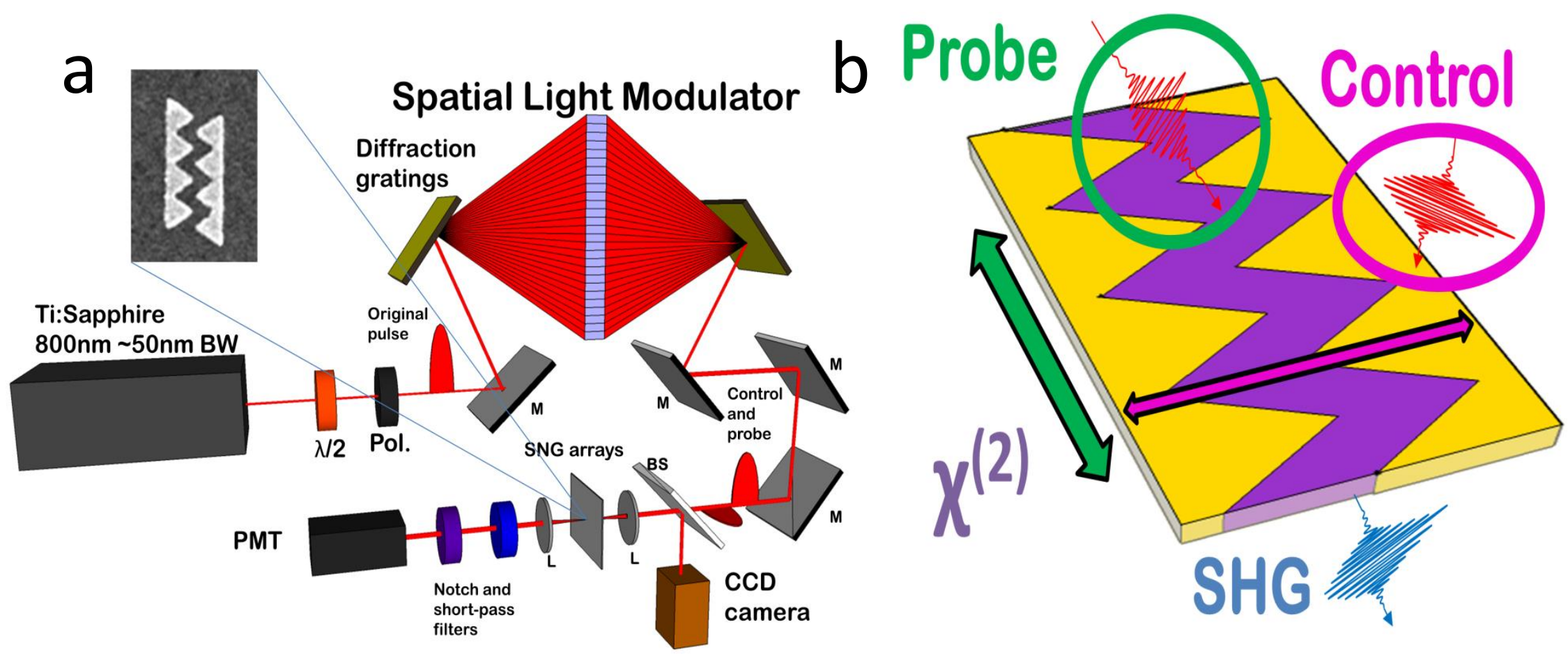}
\caption[fig2]{\label{fig:fig2}\textbf{Experimental set-up. a,} A mode-locked Ti:sapphire laser is coupled to a spatial light modulator (SLM) for control of temporal pulse shape and polarization. The SLM generates transform-limited control and probe pulses with orthogonal polarizations and variable optical power and temporal delays between each 30~fs, 800~nm pulse. \textbf{b,} The SNG structures are excited with a horizontally polarized control pulse (magenta) and probed with a vertically polarized probe pulse (green), both incident normally on the SNG array.}
\end{figure*}  

In order to demonstrate plasmonic control over the SHG signal and isolate the SHG resulting from plasmonic modulation of the dielectric, we employ a spatial light modulator (SLM) configured as shown in Fig. 2a to generate a collinear, orthogonally polarized pulse pair to simultaneously control and probe the plasmonic system. As shown in Fig. 2b, the control pulse, polarized perpendicular to the nanogap, excites the plasmon, while the second pulse probes the state of the polarized dielectric in the nanogaps. The orthogonally polarized pulses generate an interference response at the second harmonic of the input pulses that is distinctive for structures with and without PMMA. The high aspect ratio (4:1) of the SNG structure does not allow the probe pulse to be absorbed by the plasmon and thereby contribute to SHG arising from plasmon scattering.

\begin{figure*}[!tb]
\includegraphics[width=\textwidth]{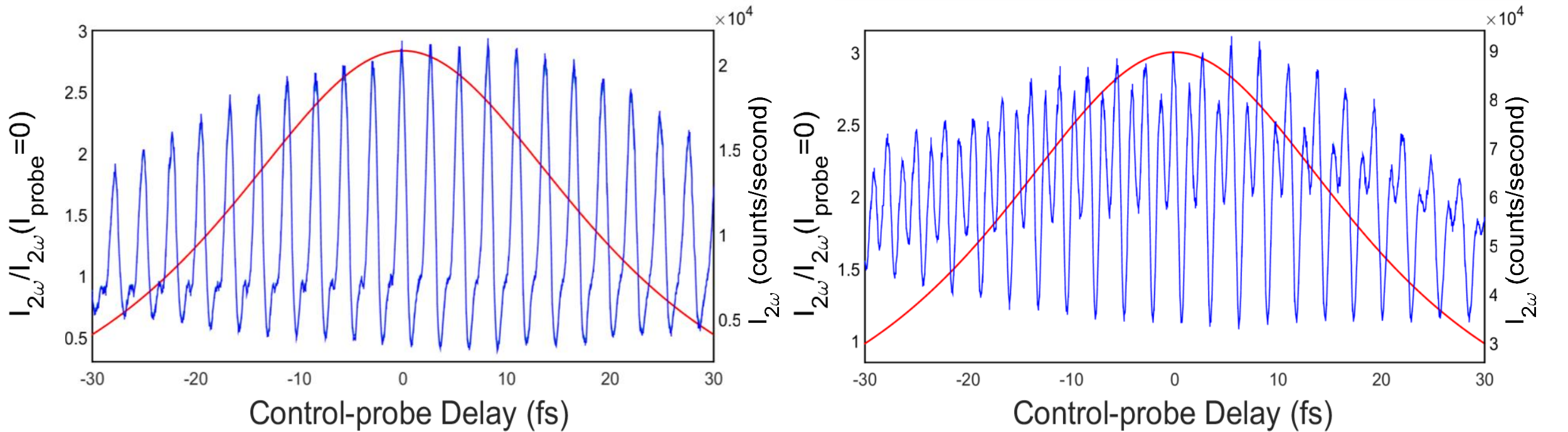}
\caption[fig3]{\label{fig:fig3}\textbf{Second-harmonic interference signal. a,} SHG as a function of control pulse-probe delay from a bare SNG array. The SHG intensity is normalized to the SHG signal from the control pulse alone. \textbf{b,} SHG from the same structure filled with the PMMA dielectric. The red solid curve in both graphs represents the temporal envelope of the control pulse.}
\end{figure*}

We demonstrate plasmonic control of the SHG output in two steps. In the first, the orthogonally polarized pulse pairs are set to equal intensity levels, with a measured average power of 10~mW in both control and probe beams. The interaction of the two beams generates characteristic SHG interference responses that differ according to whether or not there is a thin layer of PMMA in the nanogaps, as shown in Figure 3. The overall switching contrast for the SHG signal in both cases is 3:1 from peak to valley. The left- hand axes in both Figs. 3a and 3b measure the SHG intensity as a function of delay between control and probe pulses normalized to the SHG signal from the control pulse in the absence of the probe.

In Figure 3a, we see the characteristic SHG interference signature of the bare nanogap array as a function of control-probe delay time, with interference peaks spaced at time delays of $\tau$=n$\lambda$/c. As expected, the envelope of the SHG pattern extends beyond the envelope of the control pulse (red curves in Figure 3, indicating that the SHG signal is temporally broadened. The left and right-hand axes of the graphs show the normalized SHG intensity and the SHG counts for an integration time of 1~s, respectively. The imperfect linear polarization of the pulse pairs generated in the SLM contributed to the background SHG. There is a small shoulder near the intensity minima caused by plasmon-probe interference in the ITO/glass substrate and the air dielectric. Since the evanescent field of the plasmon does not extend deep into the substrate due to the absorption of the gold nanogaps and the ITO layer, this signal is relatively small. Supporting FDTD simulations can be seen in Figure S2.  

When the SNGs are filled by spin-coating the array with PMMA, on the other hand, the transmitted SHG signal from the SNG array changes dramatically, as shown in Fig. 3b. Not only does the overall SHG conversion rate, referred to the right hand axes, increase four-fold compared to the bare SNG array, but interference peaks are now observed both at delays of $\tau$=n$\lambda$/c and $\tau$=n$\lambda$/2c. Peaks at $\tau$=n$\lambda$/c correspond to constructive interference of the control and probe pulses, as before. Peaks at $\tau$=n$\lambda$/2c lie where two 800~nm waves would normally interfere destructively to create an SHG minimum; this second set of peaks results from the optical interaction of the control beam with the probe-pulse-driven dielectric polarization of the PMMA. Due to the continuous pumping of the plasmon throughout the fundamental pulse duration, the envelope of this interference pattern is the result of convolving two 30~fs pulses with the addition of the lifetime of the plasmon, $\sim$10fs.\cite{sonnichsen2002drastic} This envelope shows us that the resulting SHG pulse is approximately 70 fs in duration. A small change in the absorption of the plasmon due to the change in the dielectric environment after PMMA removal can be seen in the Supplementary Information (S1). However, the total change in absorption was less than 5\%, which is not sufficient to account for the change in SHG conversion efficiencies that follow.

For time dependent external electric fields, the second-order polarizability depends on the second order susceptibility $\chi^{(2)}$ as, 

\begin{equation}
P_{2\omega}(E)\sim\chi^{(2)}(E_{plasmon})\int_{0}^{\tau_p}[E_{probe}(t+\tau)]^2dt
\label{eq:x1}
\end{equation} 

where $E_{probe}$ represents the probe field amplitude from the pulse polarized parallel to the gap, and $E_{control}$ represents the optical field of the control pulse, polarized perpendicular to the gap, that drives the plasmon field.\cite{cornet2014terahertz} In order to properly characterize the power dependencies of the plasmon driven SHG from the dielectric, it is important to note that the PMMA dielectric is amorphous ,and therefore, either of these fields can contribute to SHG conversion.

\begin{figure*}[!tb]
	\centering
		\includegraphics[width=\textwidth]{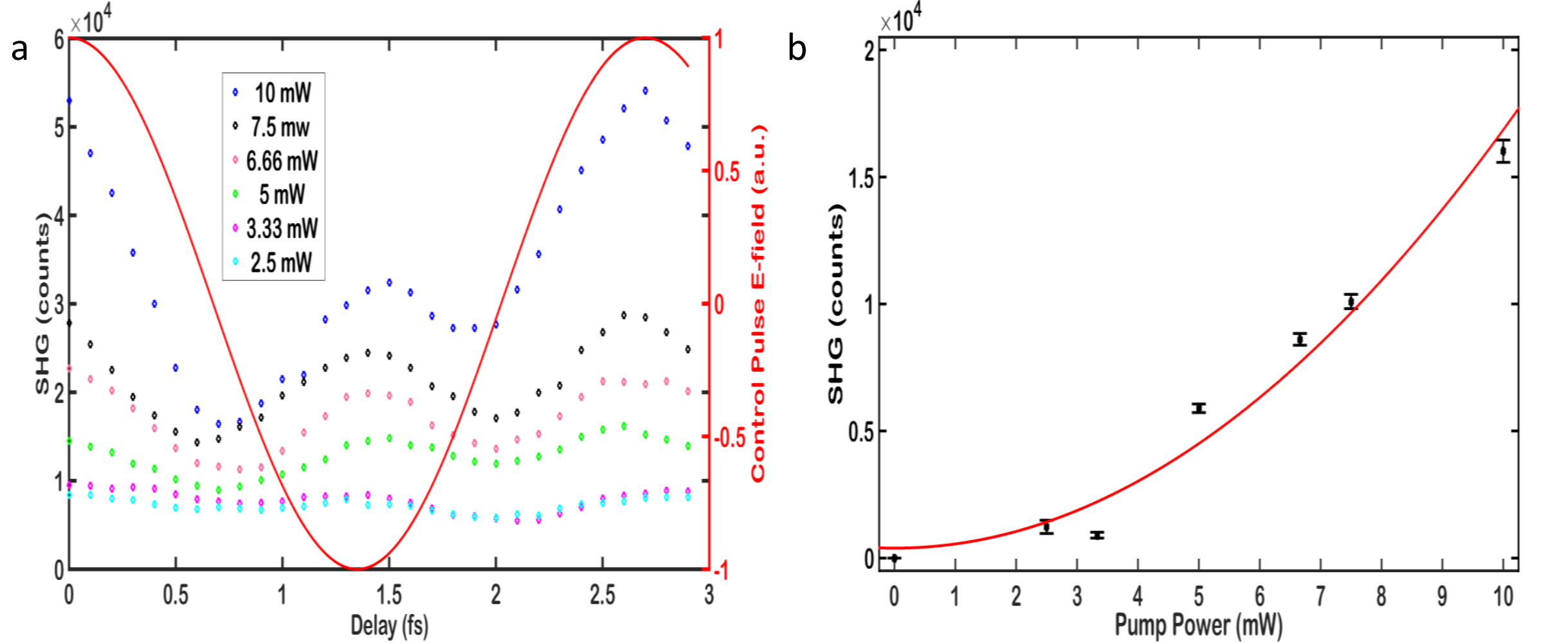}
\caption[fig4]{\label{fig:fig4}\textbf{Control pulse power dependence. a,} A single interference cycle of the SHG signal from a PMMA covered SNG for varying control pulse powers overlaid with the normalized electric field of the control pulse (red). \textbf{b,} Interference peak height from SHG maximum to minimum as a function of control pulse power. Measured interference heights are fitted to a quadratic function with no linear term with $R^2 = 0.98$.}
	\label{fig:fig4}
\end{figure*}

The second step is to demonstrate explicitly that the plasmon field drives the four-fold enhancement in SHG intensity when the gap is filled with PMMA. This is accomplished by holding the average power of the probe pulse constant at 10~mW, while raising the control-pulse power in steps from 2.5 to 10~mW. Figure 4a shows the variation in SHG yield during the first optical cycle after zero control-probe delay, normalized to the SHG yield in the absence of the probe pulse. The envelope of a single cycle of the normalized control pulse is shown in red. At 2.5~mW control-pulse power, there is already a low level of plasmon-driven SHG developing even in the minimum of the control pulse field. This rather noisy signal is close to threshold and as a result does not show an obvious coherent relationship to the control pulse.

However, for control-pulse powers at and above 5~mW, a clear oscillatory pattern in the plasmon-driven SHG signal that is nearly, but not exactly, $\pi$ radians out of phase with the control field and oscillating at 2$\omega$. The maxima in this signal occur at $\tau$=n$\lambda$/2c, as expected from the data in Fig. 3b; the minima in this signal occur very near $\tau$=n$\lambda$/4c and $\tau$=3n$\lambda$/4c. The minimum at $\tau$=3n$\lambda$/4c shows the additional interesting feature that it is not as deep, presumably because at this point the control-pulse intensity at frequency $\omega$, and with it the plasmon field, is once again increasing toward a maximum. This sub-femtosecond snapshot of of the SHG signal reveals the dynamics of the plasmon interaction with the PMMA with unprecedented temporal resolution. A possible explanation for this distinct asymmetry in the SHG signal is from the response time of the PMMA. Since the nonlinear properties of the dielectric are being altered at optical frequencies, the polarized material does not fully relax before the plasmon scattering signal associated with the $\tau$=n$\lambda$/c maxima drives a new plasmon excitation. This effect does not appear for lower control pulse powers because the magnitude of the polarization in the PMMA is not as strong as for a 10~mW control pulse. Measurements of SHG from a PMMA film of equivalent thickness showed approximately $10^{2}$~counts/second at a control pulse power of 10~mW. Changes in intensity for varying control powers were below the noise threshold for this measurement system.

Figure 4b shows the peak-to-valley amplitude of the interference intensity, measured as a function of the control pulse power. The lowest values occurring before and after the plasmonic excitation at $\tau$=n$\lambda$/2c have a difference in magnitude of $0.5*10^4$. A quadratic fit (smooth curve) indicates that $\chi^{(2)}$ depends on the square of the control field for powers well below the optical damage threshold. A model describing this second-order dependence as well as characterizing the response of highly nonlinear materials in this plasmonic geometry remains to be done.

By demonstrating that the SHG efficiency for metamaterial-controlled dielectric scales quadratically with the control field, we have provided a robust platform for ultrafast switchable metasurfaces, efficiently generating nonlinear signals with pulse durations less than 100~fs. The SNG geometry can be used to study the optical properties of thin-film materials at nanometer length scales taking advantage of the polarizing effects of electric fields oscillating at optical frequencies. The SHG spectroscopy enabled by the spatial light modulator makes it possible to separate plasmonically induced second-harmonic light within a dielectric material from light scattered by the plasmon. These experiments can be expanded to include materials with a larger $\chi^{(2)}$ as well as materials with stronger polarizabilities such as ferroelectrics. This technique is not limited to second-order nonlinearities, which makes it possible to explore third-order nonlinear effects such as phase conjugation and absorption saturation on metasurfaces.

\subsection{Materials and Methods}

Output pulses from a Ti:sapphire laser oscillator mode-locked at 83MHz (KMLabs Cascade) were passed through and compressed by a spatial light modulator (SLM) (Biophotonics Solutions Inc.) in order to achieve a transform-limited pulse 3~fs in duration at the metasurface. The laser pulse train was focused onto the SNG array with a numerical aperture objective of 0.35. The beam was then filtered by a set of two optical filters to in order to block signals from the fundamental wavelengths and processes other than SHG. The first was a high pass filter (Semrock FF01-440/SP-25) with a cutoff at 440~nm and an optical density of 0.01 at 400~nm and 5.5 at 800nmm. The second was a band pass filter (Thorlabs FB400-40) with a center wavelength at 400~nm, FWHM of 40~nm, and an optical density of 0.32 at 400~nm and 4.7 at 800nm. The resulting second harmonic signals were detected using a Hamamatsu PMT (RU-9880U-110) connected to a Photon Counter (Stanford Research Systems).

The SLM consisted of 128 liquid crystal cells. Pulse pairs were generated by creating two phase masks from alternating liquid crystal cells to create identical spectral content.\cite{pestov:09} The phase mask was then altered for each measurement to scan the probe across the control field with a resolution of 100~as. In order to reduce the power of the control pulse, separate dummy pulses were created with a vertical polarization and temporal separation from the control of greater than 200~fs. These pulses did not contribute to the SHG signal, as confirmed by a hundred-fold reduction in the SHG signal from a purely vertical excitation. This was done due to the inability to control the relative polarization and amplitudes of the pulses generated by the SLM.

The dimensions of the SNG nanostructures required for resonant operation at 800~nm were calculated using Lumerical FDTD software. The nanostructures were fabricated using standard electron-beam lithography and thermal evaporation of gold. A double-layer mask of PMMA 495A4 and 950A4 was used to optimize surface quality. All nanostructures were fabricated with a thickness of 40~nm. The SNGS were made in arrays with an inter-particle pitch of $1\mu$m in the x direction and $2\mu$m in the y-direction. A single array consisted of 82 distinct SNG particles. The SNGs were covered with 495PMMA A4 electron-beam resist to acquire the data with PMMA filling the nanogaps.

\begin{acknowledgement}
RBD, SMA and JIZ were supported by the Office of Science, U.S. Department of Energy (DE-FG02-01ER45916). AY was supported by the National Science Foundation Research Experience for Undergraduates program of the Vanderbilt Institute of Nanoscale Science and Engineering (DMR-1263182). The nanogap samples were fabricated and characterized in facilities of the Vanderbilt Institute of Nanoscale Science and Engineering, which were renovated with funds provided by the National Science Foundation under the American Re- covery and Reinvestment Act (ARI-R2 DMR-0963361). BJL and RBD acknowledge partial support from the Oak Ridge National Laboratory directed research and development program. Oak Ridge National Laboratory is operated by UT-Battelle for the U.S. Department of Energy under Contract No. DE-AC05-00OR22725.
\end{acknowledgement}

\providecommand{\latin}[1]{#1}
\providecommand*\mcitethebibliography{\thebibliography}
\csname @ifundefined\endcsname{endmcitethebibliography}
  {\let\endmcitethebibliography\endthebibliography}{}

\end{document}